\documentclass[twocolumn,amsmath,amssymb,prb,longbibliography]{revtex4-1}
\usepackage{multirow}
\usepackage{array}
\usepackage{amsmath}
\usepackage{graphicx}% Include figure files
\usepackage{dcolumn}% Align table columns on decimal point
\usepackage{bm}% bold math
\usepackage{verbatim}
\DeclareMathAlphabet \mathbfcal{OMS}{cmsy}{b}{n}
\begin{document}

\title{Anomalous Ultrafast All-Optical Hall Effect in Gapped Graphene}
\author{S. Azar Oliaei Motlagh}
%\author{Fatemeh Nematollahi}
\author{Vadym Apalkov}
\author{Mark I. Stockman}
\affiliation{Center for Nano-Optics (CeNO) and
Department of Physics and Astronomy, Georgia State
University, Atlanta, Georgia 30303, USA
}

\date{\today}
\begin{abstract}
We propose an ultrafast all-optical anomalous Hall effect in two-dimensional (2D) semiconductors of hexagonal symmetry such as gapped graphene (GG),  transition metal dichalcogenides (TMDCs), and hexagonal boron nitride (h-BN). To induce such an effect, the material is subjected to a sequence of two strong-field single-optical-cycle pulses: a chiral pump pulse followed within a few femtoseconds by a probe pulse linearly polarized in the armchair direction of the 2D lattice. Due to the effect of topological resonance, the first (pump) pulse induces a large chirality (valley polarization) in the system, while the second pulse generates a femtosecond pulse of the anomalous Hall current. The proposed effect is the fundamentally the fastest all-optical anomalous Hall effect possible in nature. It can be applied to ultrafast all-optical storage and processing of information, both classical and quantum.
\end{abstract}
%\pacs{}
\maketitle
%\section{Introduction} 

\section{Introduction}
Two-dimensional (2D) materials with honeycomb crystal structure \cite{Novoselov_et_al_Science_2016_2D_Materials_and_Heterostructures}, such as graphene, silicene, transition metal dichalcogenides (TMDCs), and hexagonal boron nitride (h-BN),  possess nontrivial topological properties in the reciprocal space \cite{Xiao_Niu_RevModPhys.82_2010_Berry_Phase_in_Electronic_Properties}. Such properties are determined by the Berry curvature, which is concentrated at the $K$ and $K^\prime$ points of the Brillouin zone. While for graphene, which is a semimetal, the Berry curvature is singular at the $K$ and $K^\prime$  points and zero elsewhere, in the gapped graphene\cite{Kjeld_et_al_PhysRevB.79.113406_2009_Gapped_Graphene_Optical_Response} (GG) and semiconductor TMDCs the Berry curvature is regular in the entire Brillouin zone with extrema at the $K$ and $K^\prime$  points. Consequently, the ultrafast electron dynamics produced by the strong optical pulses is fundamentally different in these materials\cite{Stockman_et_al_PhysRevB.93.155434_Graphene_Circular_Interferometry, Stockman_et_al_PhysRevB.98_2018_Rapid_Communication_Topological_Resonances}. 

In graphene, for a single-oscillation chiral (``circularly polarized'') pulse, the residual (left after the pulse) population of the conduction band (CB) is almost the same for the $K$ and $K^\prime$ valleys (i.e., the induced valley polarization is very weak).  For a longer pulse (with two or more optical oscillations), the valley polarization is larger; there are also pronounced fringes in the CB electron population, which form an electron interferogram caused by the accumulation of the Berry phase along the Bloch trajectories of electrons in the reciprocal space \cite{Stockman_et_al_PhysRevB.93.155434_Graphene_Circular_Interferometry}. These interferograms possess characteristic forks manifesting the presence of a quantized Berry flux of $\pm\pi$. The electron CB population distribution in the reciprocal space for both linearly \cite{Stockman_et_al_PRB_2015_Graphene_in_Ultrafast_Field} and circularly-polarized  pulses\cite{Stockman_et_al_PhysRevB.93.155434_Graphene_Circular_Interferometry} are asymmetric, which causes electric currents that have been recently  observed experimentally \cite{Higuchi_Hommelhoff_et_al_Nature_2017_Currents_in_Graphene}.

In stark contrast, the two-dimensional semiconductors (GG and TMDCs) placed in the field of chiral pulse, behave quite differently from graphene. Namely, there is a strong valley polarization induced by a circularly-polarized CW radiation of relatively low intensity \cite{Xiao_et_al_PRL_2012_Coupled_Spin_and_Valley_Physics, Zeng_et_al_Nature_Nanotechnology_2012_Valley_polarization_in_MoS2, Heinz_et_al_10.1038_Nnano.2012.96_Valley_Polarization_in_TMDC_by_Optical_Helicity, Feng_et_al_ncomms1882_2012_Valley_Selective_CD, Jones_et_al_Nature_Nanotechnology_2013_Optical_generation_of_excitonic_valley, Sie_et_al_Nature_Materials_2015_Valley-selective_optical}. 
A strong valley polarization can be introduced even by a single-oscillation 
ultrashort intense optical pulse\cite{Stockman_et_al_PhysRevB.98_2018_Rapid_Communication_Topological_Resonances}. The reason for a strong residual valley polarization in the two-dimensional semiconductors is that they have broken inversion symmetry and, consequently, a finite bandgap. As a result, for a chiral pulse that breaks the time-reversal ($\cal T$) symmetry, the valleys in the residual state of the system are populated differently. For a relatively weak CW fields, this asymmetry is due to the chiral selection rules of the transitional dipole at the $K$ and $K^\prime$ points. For an intense single-oscillation pulse, the strong valley polarization is caused by the effect of topological resonance  \cite{Stockman_et_al_PhysRevB.98_2018_Rapid_Communication_Topological_Resonances}, which is due to the interference of the topological phase (the sum of the Berry phase and the phase of the transitional dipole  matrix element) and the dynamic phase.

Ultrafast generation of a large valley polarization in the GG and TMDCs by a single-oscillation chiral pulse opens up a possibility to observe an ultrafast anomalous all-optical Hall effect. Consider a second single-cycle optical pulse that is linearly polarized in the armchair direction incident normally on the already valley-polarized solid. It is predicted to produce both a normal current in the direction of the electric field and the Hall current in the perpendicular (zigzag) direction. The latter is due to the net effect of the Berry curvature in the valley-polarized system. It changes sign for the chiral pulse of the opposite handedness. This normal current is the manifestation of the anomalous (without a magnetic field) all-optical Hall effect. The proposed all-optical anomalous Hall effect is the fundamentally fastest such an effect in nature: it takes just a single optical cycle to induce the large valley polarization and another single cycle pulse to read it out.

In this article, we consider GG, which is experimentally obtained by growing the graphene on a different substrate, i.e., on SiC  \cite{Ajayan_et_al_Review-JNN_2011_Band_Opening_in_Graphene, Conrad_et_al_PhysRevLett.115_2015_Gapped_Graphene_on_SiC}. The GG can also serve as a generic model of TMDCs. We predict the generation of an anomalous Hall current by a combination of a strong chiral pulse, which breaks the time-reversal symmetry thus playing the role of an effective magnetic field, followed by a linearly polarized probe pulse. Using the model of GG allows one to model materials with different bandgaps and to study how the anomalous Hall effect depends on the magnitude of the bandgap.

\section{Model and Main Equations}
\label{Model_and_Equations}

\subsection{Time-Dependent Schr\"odinger Equation and Its Solution}
\label{TDSE}

\begin{figure}
\begin{center}\includegraphics[width=0.47\textwidth]{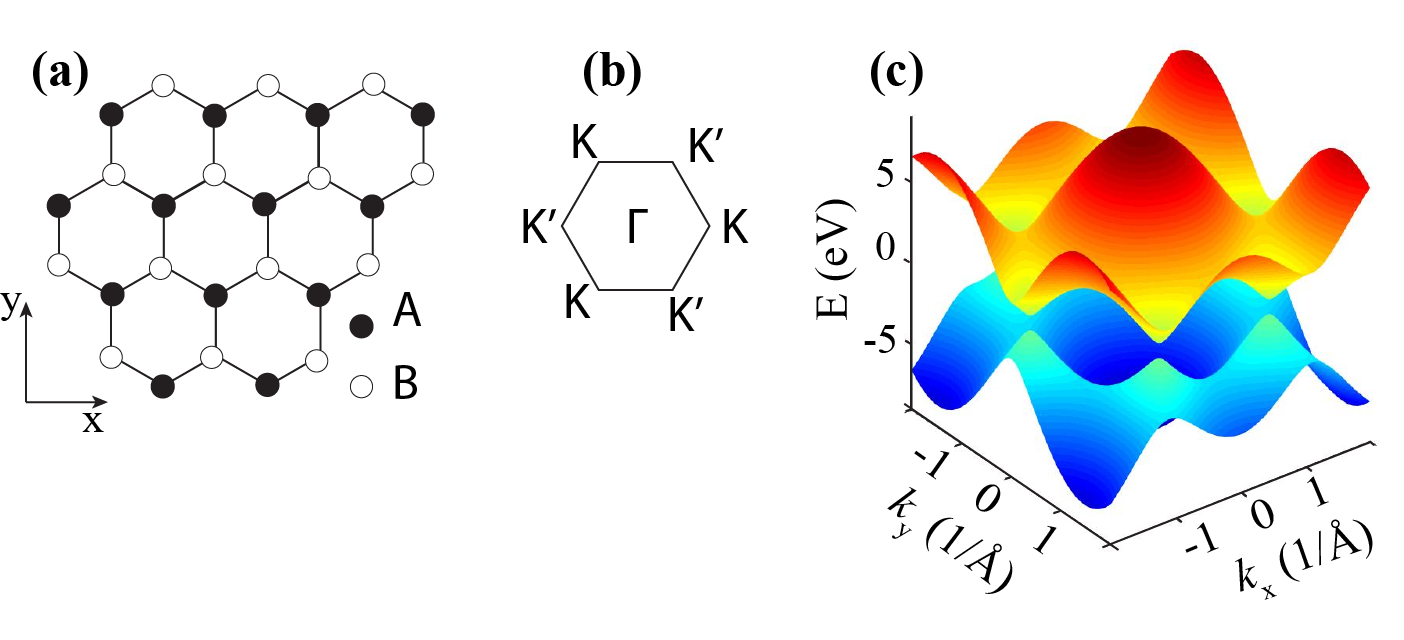}\end{center}
  \caption{(Color online) (a) Honeycomb crystal structure of graphene with sublattices A and B. (b) The first Brillouin zone of graphene with two valleys $K$ and $K^\prime$. (c) Energy dispersion a function of crystal wave vector for GG with the band gap of 1 eV.}
  \label{fig:Energy}
\end{figure}

A gap in graphene can be opened by breaking the inversion symmetry ($\cal P$), 
i.e., the symmetry between two sublattices\cite{Kjeld_et_al_PhysRevB.79.113406_2009_Gapped_Graphene_Optical_Response}, $A$ and $B$. To describe the GG, we consider two-band tight-binding Hamiltonian, which includes an extra diagonal term with the on-site energies  $\Delta_g/2$ and $-\Delta_g/2$ at two sublattices $A$ and $B$, respectively -- see Fig.\ \ref{fig:Energy}(a). This difference in the on-site energies breaks down the $\cal P$-symmetry causing the bandgaps of $\Delta_g$ to open up at the  $K$- and $K^\prime$-points -- see the schematics of the Brillouin zone in Fig.\ \ref{fig:Energy}(b). Note that the electron spectra in the $K$ and $K^\prime$ valleys are identical as protected by the time-reversal symmetry -- see  Fig.\ \ref{fig:Energy}(c), while the Berry curvatures are opposite.

Below we consider the interaction of the GG with ultrashort optical pulses 
of a few femtosecond duration. The electron scattering times in graphene and other 2D materials are on the order of or significantly longer than 10 fs -- see Refs.\ \onlinecite{Hwang_Das_Sarma_PRB_2008_Graphene_Relaxation_Time, Breusing_et_al_Ultrafast-nonequilibrium-carrier-dynamics_PRB_2011, theory_absorption_ultrafast_kinetics_graphene_PRB_2011, Ultrafast_collinear_scattering_graphene_nat_comm_2013, Gierz_Snapshots-non-equilibrium-Dirac_Nat-Material_2013, Nonequilibrium_dynamics_photoexcited_electrons_graphene_PRB_2013}. Thus, for such ultrashort optical pulses (with the duration less than 10 fs) the electron dynamics in the field of the pulse is coherent and collisionless. Consequently, it can be described by a time-dependent Schr\"odinger equation (TDSE),  
\begin{equation}
i\hbar \frac{{d\Psi }}{{dt}} = { H_{\mathbf k}(t)} \Psi~,~~~ H_\mathbf k(t) = { H}_{\mathbf k 0} - e{\mathbf{F}}(t){\bf{r}},
\label{Sch}
\end{equation}
%with Hamiltonian
%\begin{equation}
%{ H}(t) = { H}_0 - e{\bf{F}}(t){\bf{r}},
%\label{Ht}
%\end{equation}  
where $H_{\mathbf k}(t)$ is the Hamiltonian of an electron system, which consists of 
the field-free Hamiltonian, ${ H}_{\mathbf k 0}$, and the interaction Hamiltonian with the field of the pulse, $- e{\mathbf{F}}(t){\bf{r}}$. Here, $\mathbf F(t)$ is the pulse's electric field, $e$ is electron charge,  $\mathbf k$ is the electron crystal wave vector. We set $H_{\mathbf k 0}$ as the nearest-neighbor tight binding Hamiltonian for the GG \cite{Lanzara_et_al_Nat_Mat_2007_Gapped_Graphene, Kjeld_et_al_PhysRevB.79.113406_2009_Gapped_Graphene_Optical_Response, Pyatkovskiy_JPCM_2008_Plasmons_in_Gapped_Graqphene}
\begin{eqnarray}
H_{\mathbf k 0}=\left( {\begin{array}{cc}
   \Delta_g/2 & \gamma f(\mathbf k) \\
   \gamma f^\ast(\mathbf k) & -\Delta_g/2 \\
  \end{array} } \right) ,
\label{H0}
\end{eqnarray}
where $\gamma= -3.03$ eV is the hopping integral, 
\begin{equation}
f(\mathbf k)=\exp\Big(i\frac{ak_y}{\sqrt{3}}\Big )+2\exp\Big(-i\frac{ak_y}{2\sqrt{3}}\Big )\cos{\Big(\frac{ak_x}{2}\Big )},
\label{fk}
\end{equation}
and $a=2.46~\mathrm{\AA}$ is the lattice constant. 

%In the absence of electron-electron interactions and electron relaxation, the Schr\"odinger equation correctly describes the many-body electron dynamics. Note that the density matrix, which is often used for the many-body theories, in our case can be computed in terms of the solutions of Eq.\ (\ref{Sch}) as
%\begin{equation}
%\hat\rho=\sum_{\alpha\mathbf q}\left|\Psi_{\alpha\mathbf q}\right\rangle n^\mathrm{(F)}_{\alpha\mathbf q}\left\langle \Psi_{\alpha\mathbf q}\right|~,
%\label{rho}
%\end{equation}
%where $n^\mathrm{(F)}_{\alpha\mathbf q}$ is the Fermi population factor.

The energies of CB and VB are eigenvalues of $H_{\mathbf k 0}$,
\begin{equation}
E_{\alpha}(\mathbf k)=\pm\sqrt{\gamma ^2\left |{f(\mathbf k)}\right |^2+\Delta_g ^2/4}~,
\label{Energy}
\end{equation}
where signs $\pm$ are for the CB ($\alpha=c$) and the valence band (VB) ($\alpha=v$), respectively. The energy dispersion (\ref{Energy}) is shown in Fig. \ref{fig:Energy}(c).
Below we assume that initially (before the pulse) the VB is fully occupied and the CB is empty.

In solids, the applied electric field generates both the intraband (adiabatic) and interband (non-adiabatic) electron dynamics. The intraband dynamics is determined by the Bloch acceleration theorem \cite{Bloch_Z_Phys_1929_Functions_Oscillations_in_Crystals}, which 
describes the time evolution of the wave vector, $\mathbf k(t)$, in the
time-dependent electric field, $\mathbf F(t)$,
%\begin{equation}
%\hbar \frac{{d{\bf{k}}}}{{dt}} = e{\bf{F}}(t).
%\label{acceIleration}
%\end{equation}
%From this, 
\begin{equation}
{{\bf{k}}}({\bf{q}},t) = {\bf{q}} + \frac{e}{\hbar }\int_{ - \infty }^t {{\bf{F}}({t^\prime})d{t^\prime}},
\label{kvst}
\end{equation}
where $\mathbf q$ is the initial wave vector,  $\mathbf q=\mathbf k(\mathbf q,-\infty)$.

The Bloch electron trajectories of Eq. (\ref{kvst}) determine the separatrix, which is defined as a set of initial points $\mathbf q$ in the reciprocal space for which the electron trajectories pass precisely through the corresponding $K$ or $K^\prime$ points \cite{Stockman_et_al_PhysRevB.93.155434_Graphene_Circular_Interferometry}. It is a continuous line whose parametric equation is 
\begin{equation}
\mathbf q(t)=\mathbf K-\mathbf k(0,t),~~ \mathrm{or},~~  \mathbf q(t)=\mathbf K^\prime-\mathbf k(0,t)~,
\label{separatrix}
\end{equation}
where $t\in (-\infty,\infty)$ is a parameter. When the initial lattice momentum $\mathbf q$ is inside the separatrix, the corresponding Bloch trajectory, $\mathbf k(\mathbf q,t)$, encircles the $K$ or $K^\prime$ point, otherwise it leaves the $K$ or $K^\prime$ point outside.

The adiabatic solutions of Schr\"odinger equation (\ref{Sch}),  which means solutions within a single band $\alpha$ (without an interband coupling), are the well-known Houston functions \cite{Houston_PR_1940_Electron_Acceleration_in_Lattice},
\begin{equation}
 \Phi^\mathrm{(H)}_{\alpha {\bf q}}({\bf r},t)=\Psi^{(\alpha)}_{\bf{k}(\bf q,t)} ({\bf r})\exp\left(i\phi^{(\mathrm D)}_{\mathrm{\alpha}}({\bf q},t)+i\phi^{(\mathrm B)}_{\mathrm{\alpha}}({\bf q},t)\right),
\end{equation}
where $\alpha=v,c$ for the VB and CB, respectively, and $ \mathrm{\Psi^{(\alpha)}_{{\mathbf k}}} $ are the lattice-periodic Bloch functions in the absence of the pulse field. Here the dynamic phase, $\phi^\mathrm{(D)}_{\mathrm \alpha}$, and the geometric phase, $\phi^\mathrm{(B)}_{\mathrm \alpha}$, are defined as
\begin{eqnarray}
\phi^\mathrm{(D)}_{\alpha}(\mathbf q,t)= \frac{-1}{\hbar} \int_{-\infty}^t dt^\prime \left(E_\mathrm \alpha[\mathbf k (\mathbf q,t^\prime)]\right),
 \label{phi}
 \\ 
 \phi^\mathrm{(B)}_{\mathrm \alpha}(\mathbf q,t)= \frac{e}{\hbar} \int_{-\infty}^t dt^\prime \mathbf F \left(t^\prime\right)\mathbfcal{A}^{\mathrm{\alpha \alpha}}[\mathbf k (\mathbf q,t^\prime)],
 \label{phi}
\end{eqnarray}
where   $\mathbfcal{A}^{\alpha\alpha}=\left\langle \Psi^{(\alpha)}_\mathbf q  |   i\frac{\partial}{\partial\mathbf q}|\Psi^{(\alpha)}_\mathbf q   \right\rangle $ is the intraband Berry connection.
% Analytical expressions for the intraband Berry connections, $\mathbfcal{A}^{\alpha\alpha}=(\mathcal{A}^{\alpha\alpha}_x,\mathcal{A}^{\alpha\alpha}_y)$, are derived as the following expressions
%\begin{eqnarray}
%\mathcal{A}_{x}^{cc}(\mathbf k)&=&\frac{-a\gamma ^2}{\gamma ^2 |f(\mathbf k)|^2+(\Delta_g/2-E_c)^2}
% \sin \frac{3ak_y}{2\sqrt{3}}\sin{\frac{ak_x}{2}}
%\nonumber \\
 %&& =|D_x({\mathbf k})|\exp{(i\phi_{x}({\mathbf k}))},
% \label{Axcc}
%\\
%\mathcal{A}_{y}^{cc}(\mathbf k)&=&\frac{a\gamma ^2}{\sqrt{3}(\gamma ^2 |f(\mathbf k)|^2+(\Delta_g/2-E_c)^2)}\nonumber\\
% &&\times \Big(\cos{ ak_x}-\cos{\frac{\sqrt{3}ak_y}{2}}\cos{\frac{ak_x}{2}}\Big)
%\nonumber \\
 %&& =|D_x({\mathbf k})|\exp{(i\phi_{x}({\mathbf k}))},
% \label{Aycc}
% \\
%\mathcal{A}_{x}^{vv}(\mathbf k)&=&\frac{-a\gamma ^2}{\gamma ^2 |f(\mathbf k)|^2+(\Delta_g/2+E_c)^2}
% \sin \frac{3ak_y}{2\sqrt{3}}\sin{\frac{ak_x}{2}}
%\nonumber \\
 %&& =|D_x({\mathbf k})|\exp{(i\phi_{x}({\mathbf k}))},
 %\label{Axvv}
% \\
%\mathcal{A}_{y}^{vv}(\mathbf k)&=&\frac{a\gamma ^2}{\sqrt{3}(\gamma ^2 |f(\mathbf k)|^2+(\Delta_g/2+E_c)^2)}\nonumber\\
% &&\times \Big(\cos{ ak_x}-\cos{\frac{\sqrt{3}ak_y}{2}}\cos{\frac{ak_x}{2}}\Big)
%\nonumber \\
 %&& =|D_x({\mathbf k})|\exp{(i\phi_{x}({\mathbf k}))},
 %\label{Ayvv}
%\end{eqnarray}

The interband electron dynamics is determined by solutions of TDSE (\ref{Sch}).  Such solutions are parameterized by initial wave vector ${\bf q}$ and can be expanded in the basis of Houston functions $\Phi^{(H)}_{\alpha {\bf q}}({\bf r},t)$ as
\begin{equation}
\Psi_{\bf q} ({\bf r},t)=\sum_{\alpha=c,v}\beta_{\alpha{\bf q}}(t) \Phi^{(H)}_{\alpha {\bf q}}({\bf r},t),
\end{equation}
where 
$\beta_{\alpha{\bf q}}(t)$ are expansion coefficients.
%, which satisfy the following system of coupled equations
%\begin{eqnarray} &&\frac{{d{\beta _{c{\mathbf{q}}}}(t)}}{{dt}} =  - \frac{i}{\hbar }\mathbf F(t) \mathbf Q_{cv}(\mathbf q, t) \beta _{v \mathbf{q}}(t) , \nonumber \\ &&\frac{{d{\beta _{v{\bf{q}}}}(t)}}{{dt}} =  - \frac{i}{\hbar } \mathbf{F}(t)\mathbf{Q}^\ast_{cv}(\mathbf q,t)\beta _{c\mathbf{q}}(t) , \label{eq:beta_1,2} \end{eqnarray}

It is convenient to introduce the following notations
\begin{eqnarray}
\mathbfcal D^\mathrm{cv}(\mathbf q,t)&=&
\mathbfcal A^\mathrm{cv}[\mathbf k (\mathbf q,t)]\times\nonumber \\
&&\exp\left(i\phi^\mathrm{(D)}_\mathrm{cv}(\mathbf q,t)+i\phi^\mathrm{(B)}_\mathrm{cv}(\mathbf q,t)\right),
 \label{Q}
\\
\phi^\mathrm{(D)}_\mathrm{cv}(\mathbf q,t)&=&\phi^\mathrm{(D)}_\mathrm{v}(\mathbf q,t)-\phi^\mathrm{(D)}_\mathrm{c}(\mathbf q,t)
 \label{phi}
 \\ 
 \phi^\mathrm{(B)}_\mathrm{cv}(\mathbf q,t)&=&\phi^\mathrm{(B)}_\mathrm{v}(\mathbf q,t)-\phi^\mathrm{(B)}_\mathrm{c}(\mathbf q,t)\\
{\mathbfcal{A}}^\mathrm{cv}({\mathbf q})&=&
\left\langle \Psi^\mathrm{(c)}_\mathbf q  |   i\frac{\partial}{\partial\mathbf q}|\Psi^\mathrm{(v)}_\mathbf q   \right\rangle~, 
\label{D}
\end{eqnarray} 
where  ${\mathbfcal A}^{cv}(\mathbf q)$ is  the interband (non-Abelian) Berry connection \cite{Wiczek_Zee_PhysRevLett.52_1984_Nonabelian_Berry_Phase, Xiao_Niu_RevModPhys.82_2010_Berry_Phase_in_Electronic_Properties, Yang_Liu_PhysRevB.90_2014_Non-Abelian_Berry_Curvature_and_Nonlinear_Optics}, $\phi^{\mathrm{(D)}}_{cv}(\mathbf q,t)$ is the transition dynamic phase, and $\phi^{\mathrm{(B)}}_{cv}(\mathbf q,t)$ is the transition Berry phase. Note that the interband dipole matrix element, $\mathbf D^{cv}(\mathbf q)$, which determines the optical transitions between the VB and CB at a wave vector $\mathbf q$, is related to the transition Berry connection as $\mathbf D^{cv}(\mathbf q)=e \mathbfcal{A}^{cv}(\mathbf q)$.

%The non-Abelian Berry connection matrix elements can be found analytically as
%\begin{eqnarray}
%\mathcal{A}_{x}^{cv}(\mathbf k)&=&\mathcal N\Bigg(\frac{-a}{2|f(\mathbf k)|^2}\Bigg)\Bigg( \sin\frac{ak_x}{2}\sin\frac{a\sqrt{3}k_y}{2}
%\nonumber\\
%\nonumber \\&&
%&&+i \frac{\Delta_g}{2E_c}\Bigg(\cos \frac{a\sqrt{3}k_y}{2}\sin \frac{ak_x}{2}+\sin{ak_x}\Bigg)\Bigg)
%\nonumber \\
 %&& =|D_x({\mathbf k})|\exp{(i\phi_{x}({\mathbf k}))},
% \label{Ax}
%\\
%\mathcal{A}_{y}^{cv}(\mathbf k)&=&\mathcal N\Bigg(\frac{a}{2\sqrt{3}|f(\mathbf k)|^2}\Bigg)\Bigg( -1-\cos\frac{a\sqrt{3}k_y}{2}\cos\frac{ak_x}{2}
%\nonumber\\
%\nonumber \\&&
%&&+2\cos ^2 \frac{ak_x}{2}-i \frac{3\Delta_g}2{E_c}\sin \frac{a\sqrt{3}k_y}{2}\cos \frac{ak_x}{2}\Bigg)
%\nonumber \\
 %&&=|D_y({\mathbf k})|\exp{(i\phi_{y}({\mathbf k}))},
%\label{Ay}
%\end{eqnarray}
%where
%\begin{equation}
%\mathcal N=\frac{\left|\gamma f(\mathbf k)\right|}{\sqrt{\frac{\Delta_g ^2}{4}+\left|\gamma f(\mathbf k)\right|^2}}~.
%\end{equation}

With these notations, the Schr\"odinger equation in the adiabatic basis of the Houston functions (interaction representation) takes the following form
\begin{equation}
i\hbar\frac{\partial B_\mathbf q(t)}{\partial t}= H^\prime(\mathbf q,t){B_\mathbf q}(t)~,
\label{Schrodinger}
\end{equation}
where wave function (vector of state) $B_q(t)$ and Hamiltonian $ H^\prime(\mathbf q,t)$ are defined as 
\begin{eqnarray}
B_\mathbf q(t)&=&\begin{bmatrix}\beta_{c\mathbf q}(t)\\ \beta_{v\mathbf q}(t)\\ \end{bmatrix}~,\\ 
H^\prime(\mathbf q,t)&=&-e\mathbf F(t)\mathbfcal{\hat A}(\mathbf q,t)~,\\
\mathbfcal{\hat A}(\mathbf q,t)&=&\begin{bmatrix}0&\mathbfcal D^{cv}(\mathbf q,t)\\
\mathbfcal D^{vc}(\mathbf q,t)&0\\
\end{bmatrix}~.
\end{eqnarray}
Note that the interaction Hamiltonian, $H^\prime(\mathbf q,t)$, does not have the diagonal matrix elements, which is characteristic of the interaction representation. 

 We express a formal general solution of this equation in terms of the evolution operator, $\hat S(\mathbf q,t)$, as follows
 \begin{eqnarray}
 B_\mathbf q (t)&=&\hat S(\mathbf q,t)B_\mathbf q (-\infty)~,\nonumber \\
 \hat S(\mathbf q,t)&=&\hat T \exp\left[i\int_{-\infty}^t \mathbfcal{\hat A}(\mathbf q,t^\prime)d\mathbf k(t^\prime)\right]~,
 \label{S}
 \end{eqnarray}
where $\hat T$ is the well-known time-ordering operator \cite{Abrikosov_Gorkov_Dzialoshinskii_1975_Methods_of_Quantum_Field_Theory}, and the integral is affected along the Bloch trajectory [Eq.\ (\ref{kvst})]: $d\mathbf k(t)=\frac{e}{\hbar}\mathbf F(t)dt$. We solve Eq.\ (\ref{Schrodinger}) numerically for each value of the initial  reciprocal wave vector, $\mathbf{q}$. From this solution we can find the electric current, ${\mathbf J}(t) = \left\{J_x(t),J_y(t)\right\}$, generated during the pulse. 

\subsection{Current}
\label{Current}

The 4-vector electric current density is defined as $\hat j=(e\hat\rho,e\hat\rho\hat{\mathbf v})$, where $\hat\rho$ is the operator of charge density, and $\hat{\mathbf v}$ is the operator of velocity. The latter can be defined for a given lattice momentum $\mathbf k$ as 
\begin{equation}
\hat{\mathbf v}_\mathbf k=\frac{i}{\hbar}\left[H_{\mathbf k0},\mathbf r\right]~.
\label{v}
\end{equation}
This can also be identically written as
\begin{equation}
\hat{\mathbf v}_\mathbf k=\frac{1}{\hbar}\left[\frac{\partial}{\partial\mathbf k}, H_{\mathbf k0}\right]~.
\label{vk}
\end{equation}

The band-nondiagonal ($\alpha\ne \alpha^\prime$) matrix elements of the velocity can be found from Eq.\ (\ref{vk}) as
\begin{equation}
\left\langle\Psi^{(\alpha)}_\mathbf k\left\vert \hat{\mathbf v}_\mathbf k\right\vert \Psi^{(\alpha^\prime)}_\mathbf k\right\rangle=\frac{i}{\hbar}\left[E_\alpha(\mathbf k)-E_\alpha^\prime(\mathbf k)\right]\mathbfcal{A}^{\alpha\alpha^\prime}(\mathbf k)~.
\label{jab}
\end{equation}
The band-diagonal matrix element of velocity can also be obtained from Eq.\ (\ref{vk}) taking into account an identity $\left[\frac{\partial}{\partial\mathbf k},H_{\mathbf k0}\right]= \left(\frac{\partial}{\partial\mathbf k}H_{\mathbf k0}\right)$ as
\begin{equation}
\left\langle\Psi^{(\alpha)}_\mathbf k\left\vert \hat{\mathbf v}_\mathbf k\right\vert \Psi^{(\alpha)}_\mathbf k\right\rangle=\mathbf v^{(g)}_{\alpha,\mathbf k}~,
\label{vg}
\end{equation}
where $\mathbf v_{\alpha,\mathbf  k}^{(g)}=\frac{\partial}{\partial\mathbf k}E_\alpha(\mathbf k)$ is the group velocity in a band $\alpha$ at a lattice momentum $\mathbf k$.

The 2D current density in a crystal, $\mathbf J$ (called below current for brevity), is related to the electron velocity, $\mathbf v$ as $\mathbf J=\frac{e}{a^2}\mathbf v$, where $a$ is the lattice constant [see Eq.\ (\ref{fk})]. This current, $\mathbf J$, is a sum of the interband and intraband contributions, $\mathbf J(t)=\mathbf J^\text{(intra)}(t)+\mathbf J^\text{(inter)}(t)$. In accord with Eq.\ (\ref{vg}), the intraband current can be expressed as 
\begin{equation}
\mathbf J^\text{(intra)}(t)  =\frac{2e}{a^2}\sum\limits_{\alpha=\mathrm{c,v},\mathbf q}\left| \beta _{\alpha}(\mathbf q,t) \right|^2\mathbf v_{\alpha,\mathbf k(\mathbf q,t)}^{(g)}~,
\label{intra}
\end{equation}
where a factor of 2 takes into the account spin degeneracy in our model where the spin-orbit interaction is not included. Similarly, in accord with Eq.\ (\ref{jab}), the interband current is given by 
\begin{eqnarray}
  && \mathbf J^\text{(inter)}(t)=i\frac{2e}{\hbar a^2}\sum _{\substack{\mathbf q\\ \alpha,\alpha^\prime=\mathrm{v,c}\\
 \alpha\ne\alpha^\prime}}\beta _{\alpha^\prime}^\ast(\mathbf q,t)\beta _{\alpha}(\mathbf q,t)\nonumber \\&&\times\exp \{ i \phi^\mathrm{(D)}_\mathrm{\alpha^\prime\alpha}(\mathbf q,t)+ i \phi^\mathrm{(B)}_\mathrm{\alpha^\prime\alpha}(\mathbf q,t)\}\nonumber \\ 
 &&\times\left[ E_{\alpha^\prime}\left(\mathbf k(\mathbf q,t)\right)-E_\alpha \left(\mathbf k(\mathbf q,t)\right)\right] \mathbfcal A^{\alpha\alpha^\prime}\left(\mathbf k(\mathbf q,t)\right)~.
 \label{J}
\end{eqnarray} 
Note that the current is observable and, consequently, gauge-invariant despite the Berry connection being not gauge-invariant. This can be verified by using an explicit gauge transformation.

\section{Results and discussion}
\subsection{Circularly polarized pulse}
\label{Circ}

\begin{figure}
\begin{center}\includegraphics[width=0.47\textwidth]{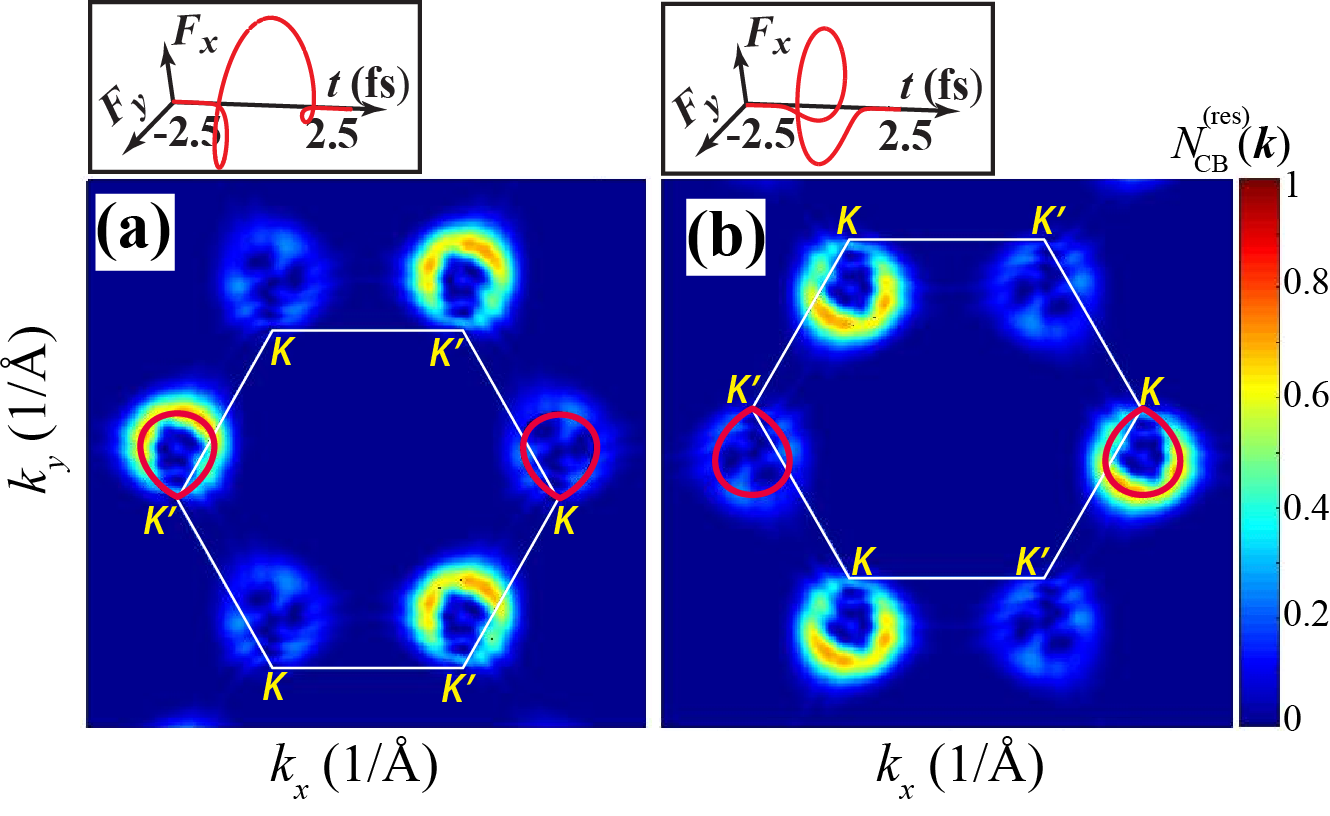}\end{center}
  \caption{(Color online) Residual CB population $N\mathrm{^{(res)}_\mathrm{CB}}(\mathbf{k})$ for GG with the band gap of $2~\mathrm{eV}$ in the extended zone picture after a chiral single-cycle excitation pulse. (a) The excitation optical pulse is left-handed with the amplitude of $F_0=0.5 ~\mathrm{V\AA^{-1}}$. Inset: Waveform of the pulse $\mathbf F(t)=\{F_x(t), F_y(t)\}$ as a function of time $t$. (b) The excitation optical pulse is right-handed with the amplitude of $F_0=0.5 ~\mathrm{V\AA^{-1}}$. Inset: Waveform of the pulse as a function of time $t$. The solid white lines show the boundary of the first Brillouin zone with the $K, K^\prime$ points indicated. The separatrix [Eq.\ (\ref{separatrix})] is shown in panels (a) and (b)  by red solid lines.
  }
  \label{fig:1RC_F0=0p5VpA_gapped_graphene}
\end{figure}

We apply an ultrafast chiral (``circularly-polarized'') optical pulse, $\mathbf F$=($F_x$, $F_y$) whose waveform is symmetric with respect to a mirror reflection in the $xz$ plane, $\mathcal P_{xz}$, as defined by the following parametrization
\begin{eqnarray}
F_x&=&F_0(1-2u^2)e^{-u^2}~,
\label{Fx}
\\
F_y&=&\pm 2F_0ue^{-u^2}~.
\label{Fy}
\end{eqnarray}
Here, $F_0$ is the amplitude of the pulse, $u=t/\tau$, where $\tau$ is a characteristic half-length of the pulse (in calculations, we choose $\tau$ = 1 fs), and $\pm$ determines the handedness: $+$  is for the
right-handed and $-$ is for the left-handed chiral (circularly polarized) pulses. In this definition,  the right-hand and left-hand pulses are $\cal T$-reversed with respect to each other. The waveforms of a right-hand pulse and a left-hand pulse are depicted in the insets in Figs. \ref{fig:1RC_F0=0p5VpA_gapped_graphene} (a) and (b), respectively.

We solve TDSE (\ref{Sch}) numerically with initial conditions $\beta_{c\mathbf q}=0$ and $\beta_{v\mathbf q}=1$, i.e., the full VB and the empty CB. An optical pulse causes interband transitions and populates the CB. After the pulse, there is  a stationary residual CB population remaining, $N\mathrm{^{(res)}_\mathrm{CB}}(\mathbf{q})=|\beta_{c\mathbf q}(t=\infty)|^2$.

For single-cycle left-handed and right-handed chiral pulses with the amplitude of 0.5 $\mathrm{V\AA^{-1}}$ the distributions of the residual CB population for  a GG with a bandgap of $\Delta_g=2~\mathrm{eV}$ are shown in Figs. \ref{fig:1RC_F0=0p5VpA_gapped_graphene}(a) and \ref{fig:1RC_F0=0p5VpA_gapped_graphene}(b), respectively. In a pristine graphene, $\Delta_g=0$, a chiral pulse  with a waveform symmetric with respect to the $\mathcal P_{xz}$ mirror reflection produces a strictly zero valley polarization \cite{Stockman_et_al_PhysRevB.100.115431_2019_Gapped_Graphene}, i.e., the $K$ and $K^\prime$ valleys are populated equally. In sharp contrast,  for a GG, there is a large valley polarization. This is due to the fact that for GG, the $\mathcal P_{xz}$ symmetry is broken. As a result, for a chiral pulse, which breaks down the time-reversal symmetry, the response of the gapped graphene in the $K$ and $K^\prime $ valleys is different. 

The ifferent populations of the $K$ and $K^\prime $ valley can be also understood from the properties of the interband coupling at two valleys. Namely, the fundamental evolution operator (\ref{S}) can be rewritten in the form
\begin{equation}
\hat S(\mathbf q,t)=\hat T\exp{\left[i\int_{-\infty}^t\mathcal{\hat A}_\Vert(\mathbf q,t^\prime)dk(t)\right]}~,
\label{S1}
\end{equation}
where the longitudinal component of the non-Abelian Berry connection is defined as $\mathcal{\hat A}_\Vert(\mathbf q, t)=\mathbfcal{\hat A}(\mathbf q,t)\mathbf F(t)/F(t)$, and $dk(t)=\frac{e}{\hbar}F(t) dt$. Explicitly, matrix $\mathcal{\hat A}_\Vert(\mathbf q,t)$ has the form
\begin{equation}
\mathcal{\hat A}_\Vert(\mathbf q,t)=\begin{bmatrix}0&\mathcal{D}^\mathrm{(cv)}_\Vert(\mathbf q,t)
\\
\mathcal{D}^{\mathrm{(cv)}\ast}_\Vert(\mathbf q,t)&0\end{bmatrix}~,
\label{Aphi}
\end{equation}
where
\begin{equation}
\mathcal{D}^\mathrm{(cv)}_\Vert(\mathbf q,t)=\left\vert\mathcal{  A}^\mathrm{(cv)}_\Vert(\mathbf k(\mathbf q,t)\right\vert \exp{\left[i\phi^\mathrm{(tot)}_\mathrm{cv}(\mathbf q,t)\right]}~,
\label{Dcv_par}
\end{equation}
and the total phase, $\phi^{(\mathrm{tot)}}_\mathrm{cv}$, is defined as
\begin{eqnarray}
 &&\phi^{(\mathrm{tot)}}_\mathrm{cv}(\mathbf q,t)=\phi^\mathrm{(D)}_\mathrm{cv}(\mathbf q,t)+\phi^\mathrm{(T)}_\mathrm{cv};
 \nonumber\\
 &&\phi^\mathrm{(T)}_\mathrm{cv}=\phi^\mathrm{(B)}_\mathrm{cv}(\mathbf q,t)+\phi^\mathrm{(A)}_\mathrm{cv}(\mathbf q,t)~.
 \label{phiQ}
 \end{eqnarray}
Here, $\phi^\mathrm{(T)}_\mathrm{cv}$ is the topological phase, and $\phi^\mathrm{(A)}_\mathrm{cv}(\mathbf q,t)=\arg{\left[\mathcal{A}_\Vert(\mathbf q,t)\right]}$ is the phase of the interband coupling amplitude. 
 
As we see from Eq.~(\ref{Dcv_par}), the interband electron dynamics is determined by the total phase $\phi^{(\mathrm{tot)}}_\mathrm{cv}$,
which is a sum of the dynamic phase, $\phi^\mathrm{(D)}_\mathrm{cv}$, and topological phase, $\phi^\mathrm{(T)}_\mathrm{cv}$. The symmetry of the dynamic and topological phases with respect to the valley index (pseudospin) is opposite: the dynamic phase is even while the topological phase is odd. Assume that in one valley, say $K$, at an initial lattice momentum $\mathbf q$, the dynamic and topological phases have opposite signs and cancel one another. This is accord with Eq.\ (\ref{S1}) will lead to a coherent accumulation of transition amplitude and, consequently, a large population of the CB.  At the same time, because of the valley antisymmetry of the topological phases, the dynamic and topological phases in valley $K^\prime$ will add to each other causing rapid oscillation of the integrand and mutual compensation of contributions over time in Eq.\ (\ref{S1}), leading to a low CB population. This is an effect of the topological resonance\cite{Stockman_et_al_PhysRevB.100.115431_2019_Gapped_Graphene}.
 
As one can see in Fig.~\ref{fig:1RC_F0=0p5VpA_gapped_graphene}(a), for the left-handed chiral pulse, the topological resonance occurs in the $K^\prime$ valley. In contrast, for the right-handed chiral pulse, it takes place in the $K$ valley [Fig.~\ref{fig:1RC_F0=0p5VpA_gapped_graphene}(b)]. Note that the conventional resonance can also be described as a cancellation of the dynamic phase $\phi^\mathrm{(D)}_\mathrm{cv}\approx \Delta_g t/\hbar $ (where $\Delta_g$ is the bandgap) and the field phase $-\omega t$, which occurs for $\omega\approx\Delta_g/\hbar$. In sharp contrast to the topological resonance, the conventional resonance  is symmetric with respect to the valley index.

\begin{figure}
\begin{center}\includegraphics[width=0.47\textwidth]{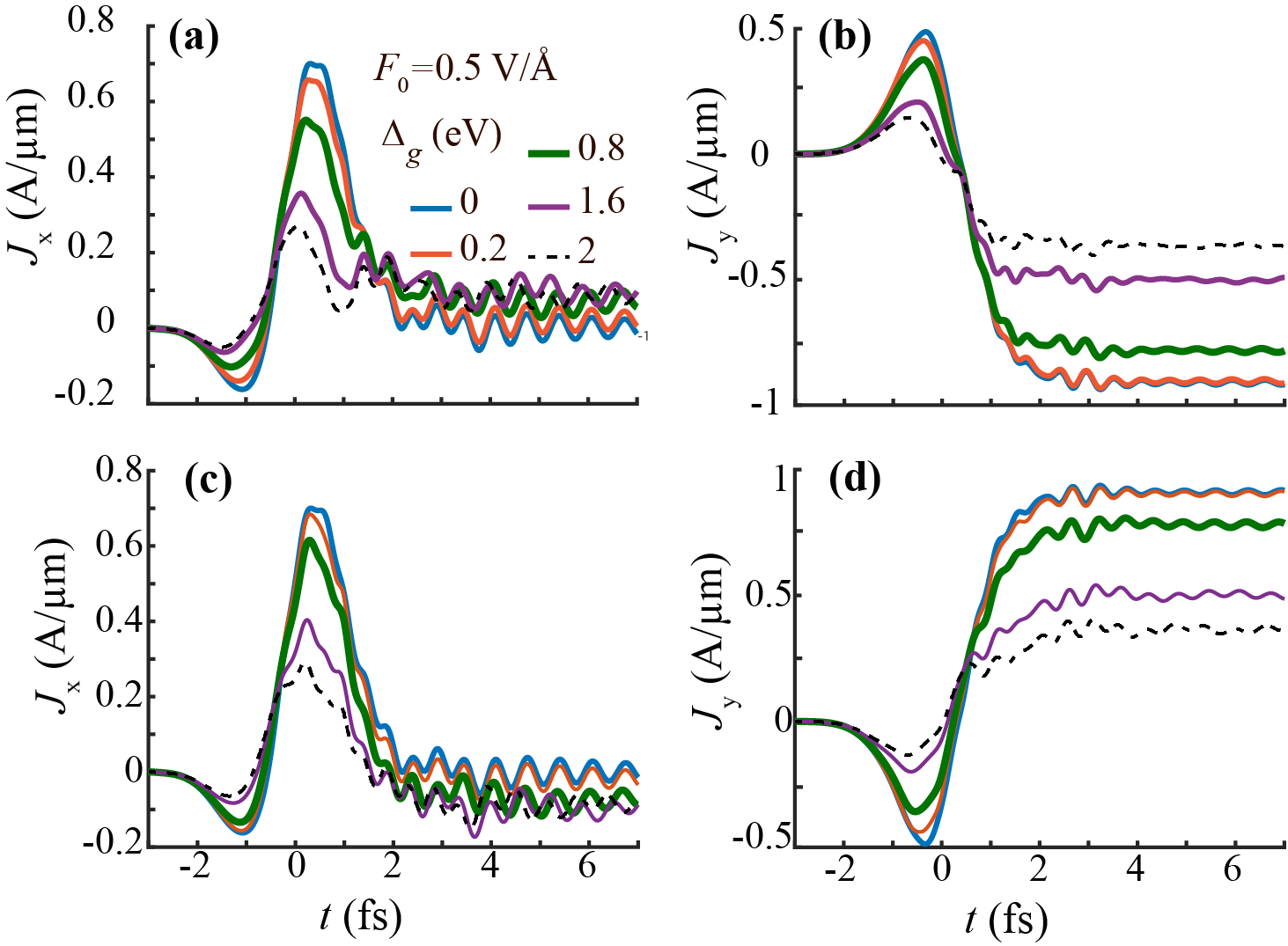}\end{center}
 \caption{(Color online) Currents $J_x$ [panel (a)] and $J_y$ [panel(b)] excited by a left-hand circularly polarized pulse with the amplitude of $F_0=0.5~\mathrm{V/\AA}$. The corresponding band gaps are marked in panel (a). In panels (c) and (d) the gapped graphene is  excited by a right-hand circularly polarized pulse. }
\label{NN_1RC_1TC_max_Fx_y_x_current_F0_0p5}
\end{figure}

The excitation pulses generate electric currents [see Eqs.\ (\ref{intra}) and (\ref{J})], which are experimentally observable -- cf.\ Ref.\ \onlinecite{Higuchi_Hommelhoff_et_al_Nature_2017_Currents_in_Graphene}. In Fig.\ \ref{NN_1RC_1TC_max_Fx_y_x_current_F0_0p5}, we show the $x$ (longitudinal, i.e., along the maximum electric field of the pulse) and $y$ (transverse)  components of the current for the left-handed and right-handed chiral pulses with an amplitude of $0.5~\mathrm{V/\AA}$ calculated for different values of the bandgap, $\Delta_g$. As one can see, both the longitudinal and transverse  currents are generated. The magnitude of these currents decrease with the bandgap along with the corresponding reduction in the CB population.  The longitudinal current, $J_x$, for graphene ($\Delta_g=0$) does not have a ballistic (dc) component: after the pulse ends, only decaying oscillations due to interband contribution are present -- see Figs.\ \ref{NN_1RC_1TC_max_Fx_y_x_current_F0_0p5} (a) and (c). This is because the dc current is purely  intraband [cf.\ Eqs.\  (\ref{intra}) and (\ref{J})] and, therefore, it is completely determined by the residual CB populations, which, for pristine graphene,  are $\mathcal P_{yz}$-symmetric due to its inherent  $\mathcal P_{xz}$ symmetry\cite{Stockman_et_al_PhysRevB.100.115431_2019_Gapped_Graphene}. This results in a complete vanishing of the ballistic $J_x$ current for pristine graphene. With the opening of the bandgap, the $\mathcal P_{xz}$ symmetry is broken, and there is a non-zero but still small ballistic current. 

\subsection{Linearly Polarized Probe and Anomalous Hall Effect}
\label{Hall_Effect}

As described above in Sec.\ \ref{Circ}, a strong single-oscillation chiral pulse creates a large valley polarization in the gapped graphene, where the carriers predominantly occupy either $K$ or $K^\prime$ valley as determined by the pulse's handedness. The resulting state has a broken $\mathcal T$ symmetry. A probe dc electric field applied to such a system will cause a Hall effect in the absence of any external or internal magnetic field, which is the anomalous  Hall effect \cite{Nagaosa_Anomalous-Hall-effect_RevModPhys_2010, Onoda_Nagaosa_PhysRevLett.90.206601_2003_Anomalous_Hall_Effect_in_2D_Ferromagnet_Metals, Hue_et_al_Science_2015_Quantum_Anomalous_Hall_Effect_in_Magnetic_TI, Cavalleri_et_al_Nat_Phys_2020_Anomalous_Hall_Effect_in_Graphene}. 

The anomalous Hall effect can be probed not only with a dc electric field but also with a linearly-polarized optical pulse applied after the strong (``pump'') chiral pulse. However, in this case to have a finite transferred charge, the linearly-polarized pulse must be strong: cf.: for a weak pulse, the total transferred charge will be zero due to the temporal averaging.

Correspondingly, we apply a nonlinear probe: a strong linearly-polarized pulse whose field is comparable to that of the chiral pulse, i.e., $\sim 0.1-0.5~\mathrm{V/\AA}$. For such a pulse, the optical nonlinearity (rectification) would define a predominant direction of the charge transfer both for longitudinal current (in the direction of the linear polarization) and for the transverse current (the anomalous Hall current). We consider a pulse linearly polarized along the $y$ axis with the following waveform 
\begin{equation}
F_x=0~,~~F_y=F_1(1-2u^2)e^{-u^2} ,
\label{LP_Pulse}
\end{equation}
where $F_1$ is the amplitude of the pulse. Note that for such a pulse in the absence of the valley polarization, there is only a longitudinal current $J_y$: a transverse current $J_x$ is forbidden by the $\mathcal P_{yz}$ symmetry of the system.

\begin{figure}
\begin{center}\includegraphics[width=0.47\textwidth]{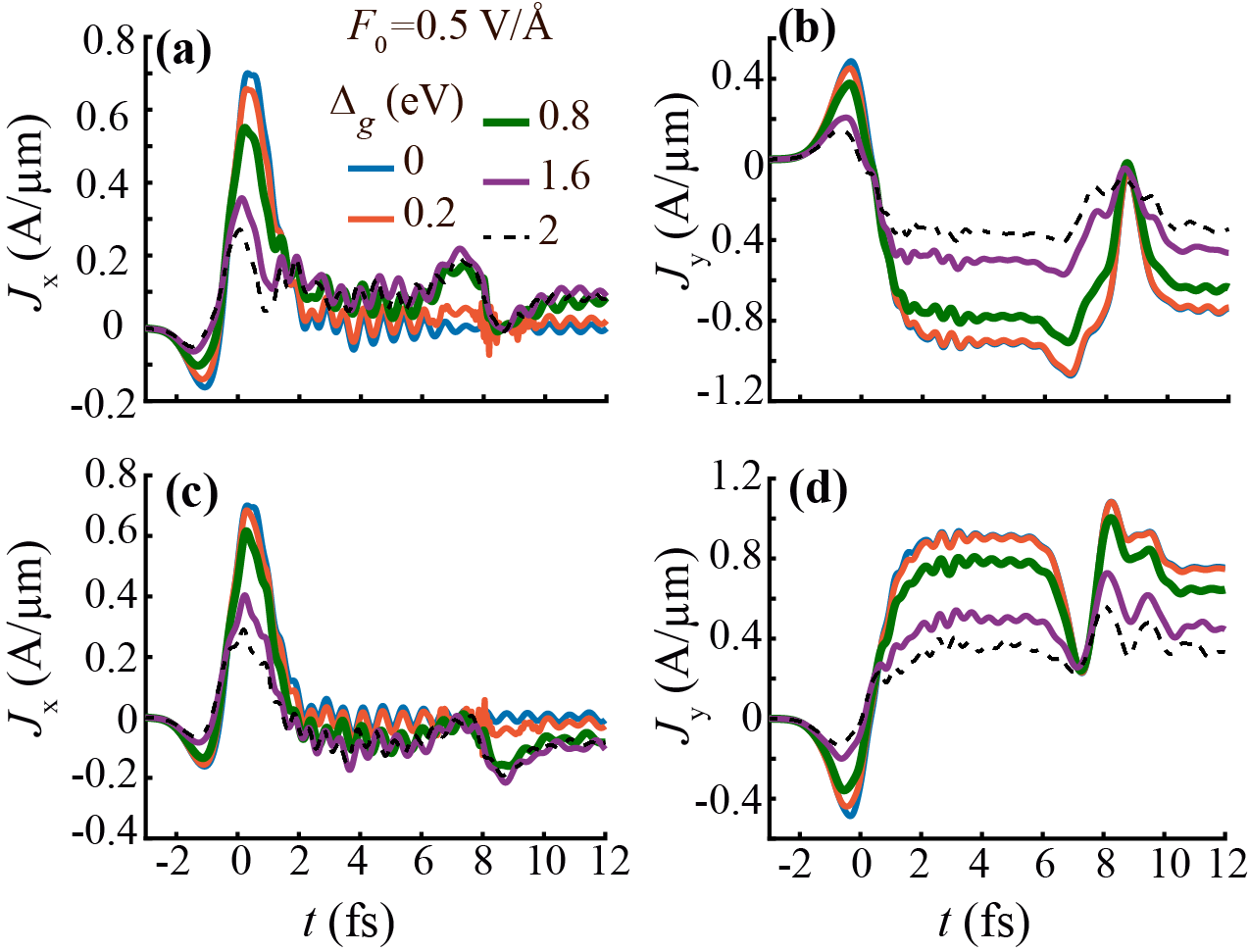}\end{center}
 \caption{(Color online). Electric currents $J_x$ [panel (a)] and $J_y$ [panel (b)] 
generated by a left-hand circularly polarized pulse ($2~\mathrm{fs}\ge t\ge -2~\mathrm{fs}$) followed by a linearly polarized pulse 
($10~\mathrm{fs}\ge t\ge 6~\mathrm{fs}$). The amplitudes of circularly and linearly polarized pulses are the same,  $F_0=F_1=0.5~~\mathrm{V/\AA}$. The corresponding band gaps are marked in panel (a).  Panels (c) and (d) are the same as panels (a) and (b) but excited by a combination of right-hand circularly polarized pulse and a linear polarized pulse. }
\label{Jx_Jy_1TC_1RC_1LY_F0c_0p5_bandgap}
\end{figure}

\begin{figure}
\begin{center}\includegraphics[width=0.47\textwidth]{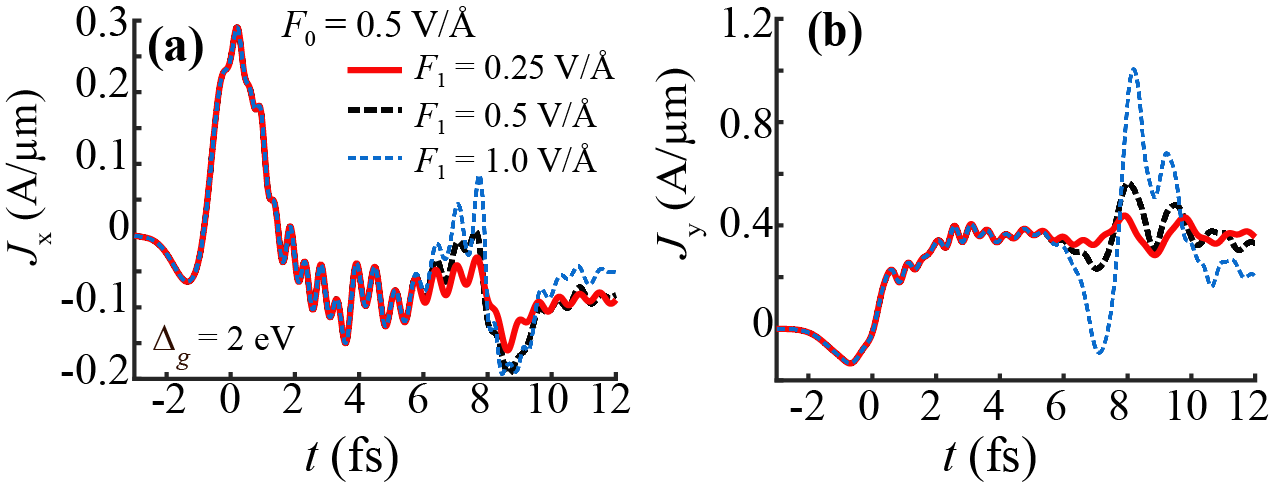}\end{center}
  \caption{(Color online). 
Electric currents $J_x$ [panel (a)] and $J_y$ [panel (b)] 
generated by a left-hand circularly polarized pulse ($2~\mathrm{fs}\ge t\ge -2~\mathrm{fs}$) followed by a linearly polarized pulse 
($10~\mathrm{fs}\ge t\ge 6~\mathrm{fs}$). The amplitude of the circularly polarized pulse is $F_0=0.5~~\mathrm{V/\AA}$, while the corresponding amplitudes of the linearly polarized pulse are marked in panel (a). The band gap is 2 eV. 
}
  \label{NN_1RC_H_max_Fx_1L_y_1RC_F0_0p5}
\end{figure}%

We apply such a linearly-polarized pulse after the chiral pulse ends (i.e., at $t\ge 6~\mathrm{fs}$). The resulting currents, which are calculated from Eqs.\ (\ref{intra}) and (\ref{J}), are shown in Fig.\ \ref{Jx_Jy_1TC_1RC_1LY_F0c_0p5_bandgap} where the strong probe pulse is applied with its center at $t=8$ fs. Note that both the longitudinal current, $J_y$ and the transverse (anomalous Hall) current $J_x$ are present in the response.

As we have already pointed out, the anomalous Hall current directly probes the valley polarization of the system. As one can see in Figs. \ \ref{Jx_Jy_1TC_1RC_1LY_F0c_0p5_bandgap} (a) and (c), for the pristine graphene ($\Delta_g=0$), the Hall current, $J_x$, is precisely zero due to the absence of the valley polarization (the corresponding lines on the graphs do not change in response to the probe pulse whatsoever).  With  the bandgap increasing, the Hall current during the linearly polarized pulse, as expected, monotonically increases [Figs. \ \ref{Jx_Jy_1TC_1RC_1LY_F0c_0p5_bandgap} (a) and (c)] because the induced valley polarization  increases with the bandgap. The anomalous Hall current, $J_x$, changes its sign with the chirality of the pump pulse as protected by the $\mathcal T$-reversal symmetry. This anomalous Hall current causes a net charge transfer in the $x$ direction, which can be measured experimentally. 

A remarkable property of the anomalous Hall current is that it has a very small ballistic component (that is the $J_x$ current after the end of the probe pulse  [Figs. \ \ref{Jx_Jy_1TC_1RC_1LY_F0c_0p5_bandgap} (a) and (c)], so it can be considered instantaneous (inertialess).  To explain this, we consider symmetry of the optical waveforms applied to the system (both chiral and linearly polarized): it is $\mathcal {TP}_{xz}$. For graphene, it is also the symmetry of the system. Thus for the graphene, the ballistic current is twice forbidden: the $\mathcal {TP}_{xz}$ symmetry forbids the valley polarization by the applied chiral pulse, and it also directly forbids the $J_x$ current because under it $J_x$ transforms to $-J_x$. For the gapped (semiconductor) materials, the $\mathcal P_{xz}$ symmetry is not exact. Nevertheless, the ballistic anomalous Hall current, $J_x$, is still very small as our computations show. Concluding, the anomalous Hall current excited by a strong linearly-polarized probe pulse acting after the strong chiral pump pulse is ultrafast (existing predominantly within the duration of the probe pulse; it is odd (changes its sign) with respect to the pump chirality.

As one can see in Figs. \ \ref{Jx_Jy_1TC_1RC_1LY_F0c_0p5_bandgap} (b) and (c), the longitudinal current, $J_y$ in response to the probe pulse monotonically decreases with the bandgap in accord with the decreasing CB population. Note that the $J_y$ current exists even for pristine graphene. Both components $J_x$ and $J_y$ generated in response to the strong probe pulse increase with its  amplitude -- see Fig.\ \ref{NN_1RC_H_max_Fx_1L_y_1RC_F0_0p5}. 

We estimate an effective Hall conductivity as $\sigma_{xy}=\Delta J_x/\Delta F_1$, where $\Delta J_x\sim 0.3~\frac{A}{\mathrm{\mu m}}$, and $\Delta F_1\sim 2~\frac{\mathrm V}{\mathrm \AA}$, where both $\Delta F_1$ and $\Delta J_x$ are obtained from Fig.\ \ref{NN_1RC_H_max_Fx_1L_y_1RC_F0_0p5} as the full range of the change of the corresponding quantity for field $F_1=1~\frac{\mathrm V}{\mathrm A}$. Using these values, we estimate the effective Hall conductivity as $\sigma_{xy}\sim 0.2 G_0$, where $G_0=\frac{e^2}{\pi\hbar}$ is the conductance quantum. 

The classical Hall conductivity is $\sigma_{xy}=enc/B$, where $n$ is the 2D electron density, $c$ is speed of light, and $B$ is the magnetic field. We may express it in terms of an effective magnetic field, $B_\mathrm{eff}$, which yields the same magnitude of $\sigma_{xy}$ as the anomalous Hall conductance, $\sim 0.2 G_0$. An estimate is  $B_\mathrm{eff}=enc/\sigma_{xy}\sim 10^9~\mathrm{G}= 10^5$ T, which is a gigantic magnetic field. Consequently, the predicted anomalous all-optical Hall effect is extraordinarily strong. It can serve as an efficient source of ultrafast currents providing a direct access to the ultrafast topological charges induced in the system.

\section{Conclusion}

A gigantic ultrafast all-optical anomalous Hall effect occurs when two strong single-oscillation optical pulses are applied to the gapped graphene or similar hexagonal-symmetry semiconductor materials such as TMDCs or h-BN. These materials possess a broken inversion symmetry and a finite direct bandgap. The two pulses, which generate the anomalous ultrafast Hall effect, are a sequence of a single-cycle chiral pulse followed by a single-cycle linearly-polarized pulse. The chiral pulse breaks down the $\mathcal T$-reversal symmetry inducing a strong valley polarization, which effectively plays the  role of an effective magnetic field. The induction of the strong valley polarization by a fundamentally fastest single oscillation chiral pulse is due to the recently predicted phenomenon of topological resonance. This is a wide-bandwidth, ultrafast effect, which is due to the mutual cancellation of the topological and dynamic phases. The topological resonance is independent of spin of electron and depends on a purely orbital dynamics of electrons in the gapped hexagonal-symmetry monolayers.

The subsequent application of a strong single-oscillation probe pulse that is linearly-polarized along the armchair edge ($y$ axis) to such a system, which acquired chirality (a large valley polarization), produces a Hall current in the zigzag direction ($x$ axis) transverse to the polarization of the probe pulse. 

The fundamental distinction and advantage of this proposed all-optical anomalous Hall effect in 2D hexagonal semiconductors from the recent proposal\cite{Sato_2019_New_J_Phys_21_093005_Theory_Optical_Induced_Anomalous_Hall_Effect_in_Graphene} and observation\cite{Cavalleri_et_al_Nat_Phys_2020_Anomalous_Hall_Effect_in_Graphene} of a light-induced anomalous Hall effect in graphene  is that ours is the fundamentally fastest anomalous effect possible in nature: it takes just a single optical period to induce the strong valley polarization and just one other optical period to read it out. Such a read out can fundamentally be done either by recording the charge transferred after the probe pulse or by observing a THz radiation emitted by the Hall current that is polarized in the $x$ direction. In sharp contrast, in Ref.\ \onlinecite{Cavalleri_et_al_Nat_Phys_2020_Anomalous_Hall_Effect_in_Graphene} the chiral excitation pulse was orders of magnitude less intense and longer: its duration was $\approx 500$ fs, i.e., in the picosecond range vs. our pulse of just $\lesssim 5$ fs duration; the read out was electrical.

There is another fundamental distinction of our predicted effect from Refs.\ \onlinecite{Sato_2019_New_J_Phys_21_093005_Theory_Optical_Induced_Anomalous_Hall_Effect_in_Graphene, Cavalleri_et_al_Nat_Phys_2020_Anomalous_Hall_Effect_in_Graphene}. Namely, a possibility to induce the strong valley polarization by a pulse with just a single optical cycle is due to the effect of topological resonance that exists only in gapped materials such as gapped graphene and 2D semiconductors (TMDCs and h-BN) but not in graphene. Therefore use of a much longer picosecond (quasi-CW) pulses in Refs.\ \onlinecite{Sato_2019_New_J_Phys_21_093005_Theory_Optical_Induced_Anomalous_Hall_Effect_in_Graphene, Cavalleri_et_al_Nat_Phys_2020_Anomalous_Hall_Effect_in_Graphene} is necessary; graphene cannot possess a single-cycle anomalous all-optical Hall effect.

The predicted ultrafast anomalous all-optical Hall effect has a potential to have applications in ultrafast memory and information processing, both classical and quantum.

\begin{acknowledgments}
Major funding was provided by Grant No. DE-SC0007043
from the Materials Sciences and Engineering Division of
the Office of the Basic Energy Sciences, Office of Science,
US Department of Energy. Numerical simulations were performed using support by Grant No. DE-FG02-01ER15213
from the Chemical Sciences, Biosciences and Geosciences
Division, Office of Basic Energy Sciences, Office of Science,
US Department of Energy. The work of V.A. was supported
by NSF EFRI NewLAW Grant No. EFMA-1741691.
\end{acknowledgments}
%\bibliography{../../BibTex/references}
%\bibliography{references}


\end{document}